\RequirePackage{fix-cm}
\documentclass[smallextended]{svjour3}       
\smartqed  
\usepackage{graphicx}
\usepackage{transparent}
\usepackage{floatrow}
\usepackage{cite}
\usepackage[colorlinks=true,breaklinks,
]{hyperref} 
\usepackage[hyphenbreaks]{breakurl} 
\usepackage{xcolor}
\definecolor{internal}{rgb}{0.6,0,0.2} 
\definecolor{citationC}{rgb}{0,0,0.6} 
\definecolor{c3}{rgb}{0.3,0,0.9} 
\hypersetup{
    linkcolor= {internal}, 
    citecolor={citationC}, 
    urlcolor={c3} 
}
\usepackage{color}
\usepackage{booktabs}
\usepackage{float}
\usepackage{amssymb}
\usepackage{amsmath}
\usepackage{graphicx}
\usepackage{subfig}
\usepackage[space]{grffile}
\usepackage{tensor}
\usepackage{amsfonts}
\usepackage{bm}
\usepackage[utf8]{inputenc}
\usepackage{slashed}
\usepackage{hyperref}
\usepackage[toc,page]{appendix}	
\usepackage{lipsum} 
\usepackage{comment}
\usepackage[percent]{overpic}
\usepackage{subfig}
\usepackage{tikz}
\usepackage{tikz}
\usetikzlibrary{arrows,chains,matrix,positioning,scopes}
\usetikzlibrary{arrows,decorations.markings}
\usetikzlibrary{tikzmark,calc}

\providecommand{\abs}[1]{\vert#1\vert}
\newtheorem{dfn}{Definition}

\def\beq{\begin{eqnarray}}
\def\eeq{\end{eqnarray}}

\definecolor{axisColor}{rgb}{0, 0, 0.5}

\begin{document}

\title{Recursion relations for gravitational lensing}


\author{Ben David Normann$^1$        \and
        Chris Clarkson$^2$
}

\institute{$^1$Faculty of Science and Technology, University of Stavanger \at 4036, Stavanger, Norway\\
              \email{bdnormann@gmail.com}           
           \and
           $^2$School of Physics \& Astronomy, Queen Mary \at
              University of London, London E1 4NS, UK,\\
              Department of Physics \& Astronomy \at University of the Western Cape, Cape Town 7535, South Africa,\\
Department of Mathematics \& Applied Mathematics, \at University of Cape Town, Cape Town 7701, South Africa.}

\date{Received: date / Accepted: date}
\maketitle
\thanks{CC was supported by STFC Consolidated Grant ST/P000592/1}\\
\begin{abstract}
The weak gravitational lensing formalism can be extended to the strong lensing regime by integrating a nonlinear version of the geodesic deviation equation. The resulting `roulette’ expansion generalises the notion of  convergence, shear and flexion to arbitrary order. The independent coefficients of this expansion are screen space gradients of the optical tidal tensor which approximates to the usual lensing potential in the weak field limit.
From lensed images, knowledge of the roulette coefficients can in principle be inverted to reconstruct the mass distribution of a lens.  In this paper, we simplify the roulette expansion and  derive a family of recursion relations between the various coefficients, generalising the Kaiser-Squires relations beyond the weak-lensing regime.
\end{abstract}
\newpage
\section{\quad Introduction}
\subsection{\quad Motivation and scope}
In 1993 Kaiser $\&$ Squires wrote an important paper \cite{kaiser93} where they found expressions for the mass in terms of the shear-field measured in galaxy clusters. Attempts had been given also previously, but these attempts involved models of the lens. Kaiser $\&$ Squires' work were further developed by Schneider, Seitz and Seitz \cite{schneider95a, schneider95b, seitz96}, and paved the way for an array of papers describing the implementation of Kaiser and Squires’ work to obtain real data. (Note, however, the recent works by Fleury et al.\cite{fleury17,fleury19a,fleury19b}, where the Kaiser-Squires theorem between shear and convergence is found to be violated due to the finiteness of the beam.). The first order correction to the weak-lensing convergence and shear has been calculated, and is called flexion \cite{goldberg05,bacon06,clarkson15}. More recently, higher order corrections have also been provided to arbitrary  order in screen-space derivatives through the so-called `Roulette' formalism \cite{clarkson16a,clarkson16b}. This is derived by integrating the fully non-linear geodesic deviation equation. 
This formalism allows one to form an extended image and calculate its extended shape from the lensing potential at the centre of the image (or more generally from the optical tidal matrix). 

The coefficients of the expansion depend on screen space derivatives of the lensing potential in the limit of a weak gravitational field, and neglecting time delay effects and so on. In this regime, we show here the formalism can be simplified in terms of the standard complex formulation of weak lensing. This then allows us to derive generalisations of the Kaiser-Squires relations to arbitrary order.

\subsection{\quad  Mathematical preliminaries}
There are two main routes to calculating the effect of a gravitational lens. One of them is to calculate all the null-geodesics converging at the point of observation. The other route is to start by the geodesic deviation equation (GDE), which typically is calculated to linear order. From a first order Taylor expansion in the Ricci rotation coefficients, the resulting equation contains one power of Riemann;
\begin{equation}
\label{GDE}
\ddot{\xi}+\tensor{R}{^a_{kbk}}\xi^b=O(\xi,\dot{\xi})^2\quad\quad\textrm{1st order GDE.}
\end{equation}
We follow the notation of \cite{clarkson15}, and the reader is referred to Figure 1 therein for an intuitive explanation of the notation: A dot ($\dot{\,\,\,}$) denotes derivative w.r.t. the null- geodesic\setcounter{footnote}{0}\footnote{Further notation  throughout: $a,b,c\cdots$denote space-time indices, $A,B,C\cdots$ are tetrad indices in screen-space. $k$ or $\xi$ as index denotes projection of that index onto $k$ or $\xi$, respectively.} and $k$ as an index represents the projection along  a congruence of null-geodesics. More accurate information about the lens is obtained by starting from the Bazanski equation \cite{bazanski77a,bazanski77b}, which takes into account terms to second order in the deviation vector $\xi$ and its derivative $\dot{\xi}$. In the context of weak lensing, this was investigated in~\cite{clarkson15}, and is the order that flexion appears. This was generalised  to arbitrary  order keeping only the maximum number of leading screen-space derivatives  in~\cite{clarkson16a,clarkson16b}\setcounter{footnote}{0}\footnote{The fully general GDE\cite{vines15}, valid to all order in all derivatives (not only screen-space), is not dealt with in this work.}, where the Roulette formalism is derived. It was  shown that if one only keeps leading order screen-space derivatives the equation to solve becomes \eqref{GDE} with the replacement
\begin{equation}
\label{repl}
\tensor{R}{^a_{k\xi k}}\quad\rightarrow\quad\sum_{n=0}^\infty\frac{1}{(n+1)!}\left(\xi^b\nabla_b\right)^n \tensor{R}{^a_{kbk}}\,.
\end{equation}
The integral solution of \eqref{GDE} with the replacement \eqref{repl} is written down by the solution procedure provided in \cite{clarkson15}. The resulting integral solution is then expanded into maps for each order $m$ of screen-space derivatives. It is shown how the general solution of the equation may be written as a sum over modes called `roulettes'. The derivation is rather involved, and we refer the reader to \cite{clarkson16a} for a comprehensive overview, and \cite{clarkson16b} for the detailed derivation. The general $m$'th order map is given as Eqs. (11)-(14) in \cite{clarkson16a}. The completely general roulette amplitudes $\alpha^m_s\,,\,\beta^m_s$ (even modes) and $\bar{\alpha}^m_s\,,\,\bar{\beta}^m_s$ (odd modes) are given as Eqs. (89)-(94) in \cite{clarkson16b}. With a symmetric amplification matrix, such as in the flat-sky approximation, all the odd modes vanish. In the present work we restrict attention to the weak-field approximation and neglect all but the highest number of screen space derivatives of the  otential. Linearising around Minkowski space we write the perturbations with respect to Poisson gauge as
\begin{equation}
    \label{pert}
    {\rm d}s^2=-(1+2\Phi){\rm d}\eta^2+(1-2\Psi)\gamma_{ij}{\rm d}x^i{\rm}x^j,
\end{equation}
and define a lensing potential
\begin{equation}
\label{lensPot}
    \psi=\int_0^\chi{\rm d}\chi'\left(\frac{\chi-\chi'}{\chi \chi'}\right)(\Phi+\Psi).
\end{equation}
We refer the reader to the discussion around eq. 102 in \cite{clarkson16b} for further details.

\subsection{\quad  Structure of the paper}
The rest of the paper is structured as follows. In Section \ref{sec:RouletteModes}
we first write down recursion relations relating the roulette amplitudes to each other. Thereafter, the Roulette formalism (in the above given approximations) is cast into a compact form, using complex notation. We use this to write down a very neat expression for the resulting complex spin roulette amplitudes $\gamma^m_s=\alpha^m_s+\beta^m_s$ that invoke no sums or integrals. In Section \ref{Sec:Recursion} we formulate the recursion relations of the previous section in complex notation. In doing so, it becomes natural to introduce operators, and to write down operator identities. The action of the operator is to take in any one (complex) roulette amplitude and transform it to any other. Then, in Section \ref{sec:Comp} we compare our results with that of earlier works, and confirm agreement. In Section \ref{sec:derivatives} we use the recursion relations to find general expressions for the lensing potential derivatives in terms of the spin roulette amplitudes $\alpha^m_s$ and $\beta^m_s$. Finally, in Section \ref{sec:concl}, we conclude.


\section{\quad Roulette modes in the weak-field approximation}
\label{sec:RouletteModes}
For a general \textit{thin lens} in the \textit{weak-field} and \textit{flat-sky} approximation the Roulette amplitudes \cite{clarkson16b} take the form 
\begin{eqnarray}
&\alpha_s^m=-2^{-\delta_{0s}} \chi^{m+1}\sum_{k=0}^m{m\choose k}\left(\mathcal{C}_s^{m(k)}\partial_{\rm X}+\mathcal{C}_s^{m(k+1)}\partial_{\rm Y}\right)\partial_{\rm X}^{m-k}\partial_{\rm Y}^k\psi\label{Alpha},\\
&\beta_s^m=-\chi^{m+1}\sum_{k=0}^m {m\choose k}\left(\mathcal{S}_s^{m(k)}\partial_{\rm X}+\mathcal{S}_s^{m(k+1)}\partial_{\rm Y}\right)\partial_{\rm X}^{m-k}\partial_{\rm Y}^k\psi\label{Beta},
\end{eqnarray}
where $\delta$ is the ordinary Kronecker delta\footnote{Hence; $\delta_{0s}=1$ if and only if $s=0$. The reason for this extra factor of $1/2$ for the spin-0 modes comes from the fact that Eq.110-111 in \cite{clarkson16b} are valid only for $s\,>\,0$. The case $s=0$ must be derived from Eq. 93 therein, and results in an extra factor of $1/2$. This factor is not needed in the expression for $\beta^m_s$, since these coefficients vanish for $s=0$ anyway.} and $\chi$ is the distance from the observer to the lens. $X,Y$ are coordinates in the lens-plane and $\psi\,=\,\psi(X,Y)$ is the lensing potential given by Eq.\,\eqref{lensPot}. Also the spin $s$ is restricted sucht that $0\,\leq\,s\,\leq\,m+1$ and the roulette amplitudes $\alpha^m_s\,,\,\beta^m_s$ may be non-zero only if $m+s$ is odd. Finally;
\begin{eqnarray}
&\mathcal{C}_s^{m(k)}=\frac{1}{\pi}\int_{-\pi}^{\pi}{\rm d}\theta\sin^k\theta\cos^{m-k+1}\theta\cos s\theta\label{C},\\
&\mathcal{S}_s^{m(k)}=\frac{1}{\pi}\int_{-\pi}^{\pi}{\rm d}\theta\sin^k\theta\cos^{m-k+1}\theta\sin s\theta.\label{S}
\end{eqnarray}
The equations \eqref{Alpha} and \eqref{Beta} are horrendous, but it is easy enough to calculate $\alpha_1^0$ and $\beta_1^0$. We find
\begin{equation}
\alpha^0_1=-\chi\partial_X\psi\quad\quad\quad\textrm{and}\quad\quad\quad\beta^0_1=-\chi\partial_Y\psi.\label{lowestmode}
\end{equation}
By the recursion relations introduced below, these two `lowest modes' of expansion will actually suffice. The recursion relations we find are:
\begin{eqnarray}
&\alpha_{s+1}^{m+1}=(C_+^+)_{s+1}^{m+1}\left(\partial_{\rm X}\alpha_s^m-\partial_{\rm Y}\beta_s^m\right)\label{rec1},\\
&\beta_{s+1}^{m+1}=(C_+^+)_{s+1}^{m+1}\left(\partial_{\rm X}\beta_s^m+\partial_{\rm Y}\alpha_s^m\right)\label{rec2},
\end{eqnarray}
and
\begin{eqnarray}
&\alpha_{s-1}^{m+1}=(C_-^+)_{s-1}^{m+1}\left(\partial_{\rm X}\alpha_s^m+\partial_{\rm Y}\beta_s^m\right)\label{rec3},\\
&\beta_{s-1}^{m+1}=(C_-^+)_{s-1}^{m+1}\left(\partial_{\rm X}\beta_s^m-\partial_{\rm Y}\alpha_s^m\right).\label{rec4}
\end{eqnarray}
Here $(C_+^+)$ and $(C_-^+)$ are algebraic coefficients
given by
\begin{equation}
(C_+^+)_s^m=2^{\delta_{0(s-1)}}\frac{m+1}{m+1+s}\chi\quad\quad\quad\textrm{and}\quad\quad\quad (C_-^+)_s^m=2^{-\delta_{0s}}\frac{m+1}{m+1-s}\chi\label{AandB}.
\end{equation}
The signs in the name of the coefficients $(C_+^+)$ and $(C_+^-)$ are there to indicate the direction of change in the value $m$ (superscript) and $s$ (subscript).
We provide a more compact notation for these relations (and the inverses) in Section \ref{Sec:Recursion}. In order to do so, however, we must first introduce complex variables.

\subsection{\quad  Complex notation}
Define first the complex variables
\begin{equation}
\gamma_s^m=\alpha_s^m+i\beta_s^m.
\end{equation}
By use of \textit{Pascal's triangle} we now, upon a bit of straight forward algebra, obtain from equations \eqref{Alpha} and \eqref{Beta} the expression
\begin{eqnarray}
\gamma_s^{m}=-2^{-\delta_{0s}}\chi^{m+1}\sum_{k=0}^{m+1}{m+1\choose k}\mathcal{D}_s^{m(k)}\partial_{\rm X}^{m+1-k}\partial_{\rm Y}^k\psi,\label{Gamma2}
\end{eqnarray}
where we have defined
\begin{eqnarray}
&\mathcal{D}_s^{m(k)}=\mathcal{C}_s^{m(k)}+i\mathcal{S}_s^{m(k)}=\frac{1}{\pi}\int_{-\pi}^{\pi}{\rm d}\theta\sin^k\theta\cos^{m-k+1}\theta\,{\rm e}^{i\cdot s\theta}\label{D}.
\end{eqnarray}
Define similarly the differentiation operator
\begin{equation}
\partial_c=\partial_X+i\partial_Y.
\end{equation}
Furthermore, let $^*$ denote the complex conjugate, and define also $\Box=\partial_X^2+\partial_Y^2$. Then 
\begin{equation}
\label{Box}
\Box=\partial_c\partial_c^*.
\end{equation}
To be explicit, the operator inverse $\Box^{-1}$ is defined such that $(\Box\circ\Box^{-1})f=(\Box^{-1}\circ\Box)f=f$. Here $f$ is a function. A particularly useful expression for $\gamma_s^m$ may be found by use of the recursion relations written down above (and complexified in the next section). To keep the algebra in the main text at a minimum we only quote the result,
\begin{equation}
\label{GamNice1}
\gamma_s^m=\Gamma_s^m\Box^{a^{-}}\partial_c^s\psi,
\end{equation}
and refer the reader to Appendix \ref{App:DerOfPot} for a derivation of this relation. Here $ a^-=(m+1-s)/2$ and $\Gamma_s^m$ are numerical coefficients given by
\begin{eqnarray}
\label{matrix}
\Gamma_s^m=
\begin{cases}
-(2^{-\delta_{0s}})\frac{\chi^{m+1}}{2^m}{m+1\choose a^-}\quad \quad\quad m+s\quad\textrm{odd,}\\
\quad\quad 0\quad\quad\quad\quad\quad\quad\quad\quad\quad\quad\quad\phantom{0}\textrm{else}.
\end{cases}
\end{eqnarray}
Note that the expression for $\gamma_s^m$ given by \eqref{GamNice1} does not involve any sum or integral. Hence it is computationally much more economic then the foregoing expression \eqref{Gamma2}, which involve a sum over integrals. Below we have written down the first few complex roulette amplitudes by use of equation \eqref{GamNice1}. Mode $\gamma_s^m$ is given in row $s$, column $m$, starting with $s=0,m=0$ in the upper left corner. 
\begin{equation}
\left(
\begin{array}{ccccc}
 0 & -\frac{1}{2}\chi ^2 \Box\psi  & 0 & -\frac{3}{8} \chi ^4 \Box^2\psi \\
 -\chi  \partial_c\psi & 0 & -\frac{3}{4} \chi ^3 \Box\partial_c\psi & 0 \\
 0 & -\frac{1}{2} \chi ^2 \partial_c^2\psi & 0 & -\frac{1}{2} \chi ^4 \Box\partial_c^2\psi \\
 0 & 0 & -\frac{1}{4} \chi^3 \partial_c^3\psi & 0\\
 0 & 0 & 0 & -\frac{1}{8} \chi ^4 \partial_c^4\psi\\
\end{array}
\right)
\end{equation}

\section{\quad \textit{D}-operators: introducing a simpler notation}
\label{Sec:Recursion}
We define in this section a set of raising and lowering operators that will serve as recursion relations (on operator form) between the different amplitudes. This language is very natural, and allows for computing amplitudes without invoking the whole mathematical machinery of the underlying theory. This is highly beneficial, since the theory is rather involved, even with the approximations adopted in this work (cf. the sums and integrals in  Eqs. \eqref{Alpha}-\eqref{S}). In want of a better name we call the identity operator $D$,\setcounter{footnote}{0}\footnote{The reason for the root name $D$ is that these operators will become differentiation operators.} and define 
\begin{equation}
\begin{split}
& D\gamma=\gamma\quad\quad\quad\quad\textit{identity element},\\
& D^+\gamma=D^-\gamma=D_+\gamma=D_-\gamma=0.
\end{split}
\end{equation}
The connection to the Roulette formalism is made by defining two basis-operators $D^+_+$ and $D^+_-$ from which all other operators may be constructed. We define
\begin{eqnarray}
\label{BaseOp}
\textit{basis-operators}\phantom{0000000000}
\begin{cases}
D^+_+=C_+^+\partial_c,\\
D^+_-=C_-^+\partial_c^*.
\end{cases}
\end{eqnarray}
Here $C_+^+$ and $C_-^+$ are numerical factors given by \eqref{AandB}. Note that $C_+^+$ and $C_-^+$ will depend on the particular value of $m$ and $s$. The recursion relations are now given by applying the basis-operators to $\gamma_s^m$, requiring that they act such that $D^+_+:\gamma^m_s\,\to\,\gamma^{m+1}_{s+1}$ and $D^+_-:\gamma^m_s\,\to\,\gamma^{m+1}_{s-1}$ for all valid ($m,s$)-positions. In particular,
\begin{eqnarray}
\label{gammaRel}
\textbf{Recursion relations}\phantom{0000000000}\begin{cases}
\gamma^{m+1}_{s+1}=D^+_+\gamma^m_s=(C_+^+)^{m+1}_{s+1}\partial_c\gamma^m_s,\\
\gamma^{m+1}_{s-1}=D^+_-\gamma^m_s=(C_-^+)^{m+1}_{s-1}\partial_c^*\gamma^m_s.
\end{cases}
\end{eqnarray}
The $+$ and $-$ signs are hence there to indicate whether we add or subtract to the number $m$ (upper index) and $s$ (lower index). Figure \ref{Fig1} illustrates how the operators propagate an amplitude to one of its neighbours. The above recursion relations are the same as the relations given in the previous section, Eqs. \eqref{rec1}-\eqref{rec4}. Note that \textbf{the coefficients are always evaluated at the end-point position} (see the example below).

A proof of the recursion relations can be found in Appendix \ref{App:A}. The two recursion relations are independent of each other and all non-zero modes $\alpha_s^m$ and $\beta_s^m$ can therefore be obtained by these relations by starting from the lowest mode\footnote{Or any other, for that sake!} $\gamma^0_1=\alpha^0_1+i\beta^0_1$, which is known from Eq. \eqref{lowestmode}.

We shall note in passing that any higher order operator $D_{b+}^{a +}$ such that $\gamma_{s+b}^{m+a}=D_{b+}^{a +}\gamma_s^m$ may be constructed from the basis-operators. We refer the interested reader to  Appendix \ref{app:groups} for an explanation and a more detailed scrutiny of these operators. Here we continue instead with an example, to show that these operators work in a very intuitive manner. To illustrate, let us calculate $\gamma_2^3$ from $\gamma^0_1$. 
\begin{equation}
\begin{split}
\gamma_2^3&\,=\,\gamma_{1+1}^{0+3}\,=\,D_{+}^{3+}\gamma_1^0\,=\,(D_-^+)\circ D_{2+}^{2+}\gamma_1^0 \,=\,(D_-^+)\circ(D_+^+)^{2}\gamma_1^0 \\
&\,=\,(C_-^+)_2^3(C_+^+)_3^2(C_+^+)_2^1\Box\partial_{\rm c}^2\gamma_1^0\,=\, \frac{1}{2}\chi^3\Box\partial_{\rm c}\gamma_1^0\,=\, -\frac{1}{2}\chi^4\Box\partial_{\rm c}^2\psi.
\end{split}
\end{equation}
In the last equality we inserted for $\gamma_1^0$ from \eqref{GamNice1}, just to show that we obtain the correct expression, as compared with $\gamma^3_2$ in  \eqref{matrix}). And again: Note that the coefficients are all calculated at the end-point of each step.

\subsection{\quad  \quad Inverse operators} 
\begin{figure}[h]
\hspace*{-1.5cm} 
\begin{tikzpicture}
\draw [dashed,decoration={markings,mark=at position 1 with
    {\arrow[scale=2,>=stealth]{>}}},postaction={decorate}]  (0,0) -- (7.8,0);
\draw [dashed,decoration={markings,mark=at position 0 with
    {\arrow[scale=2,>=stealth]{<}}},postaction={decorate}] (-7.6, 0) -- (0,0);
\draw[dashed] (0,0) -- (0,8);
\draw[dashed]  (0, 0) -- (0,-8);
\draw [decoration={markings,mark=at position 1 with
    {\arrow[scale=2,>=stealth]{>}}},postaction={decorate}] (0,0) -- (3.9,3.9);
\draw [decoration={markings,mark=at position 1 with
    {\arrow[scale=2,>=stealth]{>}}},postaction={decorate}] (0,0) -- (-3.9,-3.9);
\draw [decoration={markings,mark=at position 1 with
    {\arrow[scale=2,>=stealth]{>}}},postaction={decorate}] (0,0) -- (3.9,-3.9);
\draw [decoration={markings,mark=at position 1 with
    {\arrow[scale=2,>=stealth]{>}}},postaction={decorate}] (0,0) -- (-3.9,3.9);
\draw [dashed] (4,4) -- (8,0);
\draw [dashed] (4,4) -- (0,8);
\draw [dashed,decoration={markings,mark=at position 1 with
    {\arrow[scale=2,>=stealth]{>}}},postaction={decorate}] (4,4) -- (7.2,0.8);
\draw [dashed] (-4,4) -- (0,8);
\draw [dashed,decoration={markings,mark=at position 1 with
    {\arrow[scale=2,>=stealth]{>}}},postaction={decorate}] (-4,4) -- (-7.2,0.8);
\draw [dashed] (-8,0) -- (-4,-4);
\draw [dashed] (-4,-4) -- (0,-8);
\draw [dashed] (4,-4) -- (8,0);
\draw [dashed] (4,-4) -- (0,-8);
\node[blue,fill=white] at (0,0.3) {$(m,s)$};
\draw[fill] (0,0) circle [radius=0.1];
\node[blue,fill=white] at (4,4.3){$(m+1,s-1)$};
\draw[fill] (4,4) circle [radius=0.1];
\node[blue,fill=white] at (8,0.43){$(m+2,s)$};
\draw[fill] (8,0) circle [radius=0.1];
\node[blue,fill=white] at (4,-4.35){$(m+2,s)$};
\draw[fill] (4,-4) circle [radius=0.1];
\node[blue,fill=white] at (0,-8.35){$(m,s+2)$};
\draw[fill] (0,-8) circle [radius=0.1];
\node[blue,fill=white] at (-4,-4.35){$(m-1,s+1)$};
\draw[fill] (-4,-4) circle [radius=0.1];
\node[blue,fill=white] at (-8,0.43){$(m-2,s)$};
\draw[fill] (-8,0) circle [radius=0.1];
\node[blue,fill=white] at (-4,4.35){$(m-1,s-1)$};
\draw[fill] (-4,4) circle [radius=0.1];
\node[blue,fill=white] at (0,8.35){$(m,s-2)$};
\draw[fill] (0,8) circle [radius=0.1];
\node[black,fill=white] at (2,2){$D_-^+$};
\node[black,fill=white] at (2,-2){$D_+^+$};
\node[black,fill=white] at (-2,2){$D_-^-$};
\node[black,fill=white] at (-2,-2){$D_+^-$};
\node[black,fill=white] at (4,0){$D^{++}$};
\node[black,fill=white] at (-4,0){$D^{--}$};
\node[black,fill=white] at (6,2){$D_+^+$};
\node[black,fill=white] at (-6,2){$D_+^-$};
\end{tikzpicture}
\caption{The figure illustrates how the $D$-operators propagate the amplitude at a position $(m,s)$ to one of its neighbours. Solid lines show the basis-operators $D_+^+$ and $D_-^+$ and their inverses ($D_-^-$ and $D_+^-$) acting at $(m,s)$. Stippled lines connect other points. Some of them are shown with $D$-operators. Note that for any $D$-operator acting from $(m,s)\rightarrow(m+M,s+S)$ there corresponds an inverse $D$-operator acting from $(m+M,s+S)\rightarrow(m,s)$. This is not shown in the Figure. Higher order operators will connect amplitudes separated by a larger distance in the $(m,s)$-configuration-space.}\label{Fig1}
\end{figure}
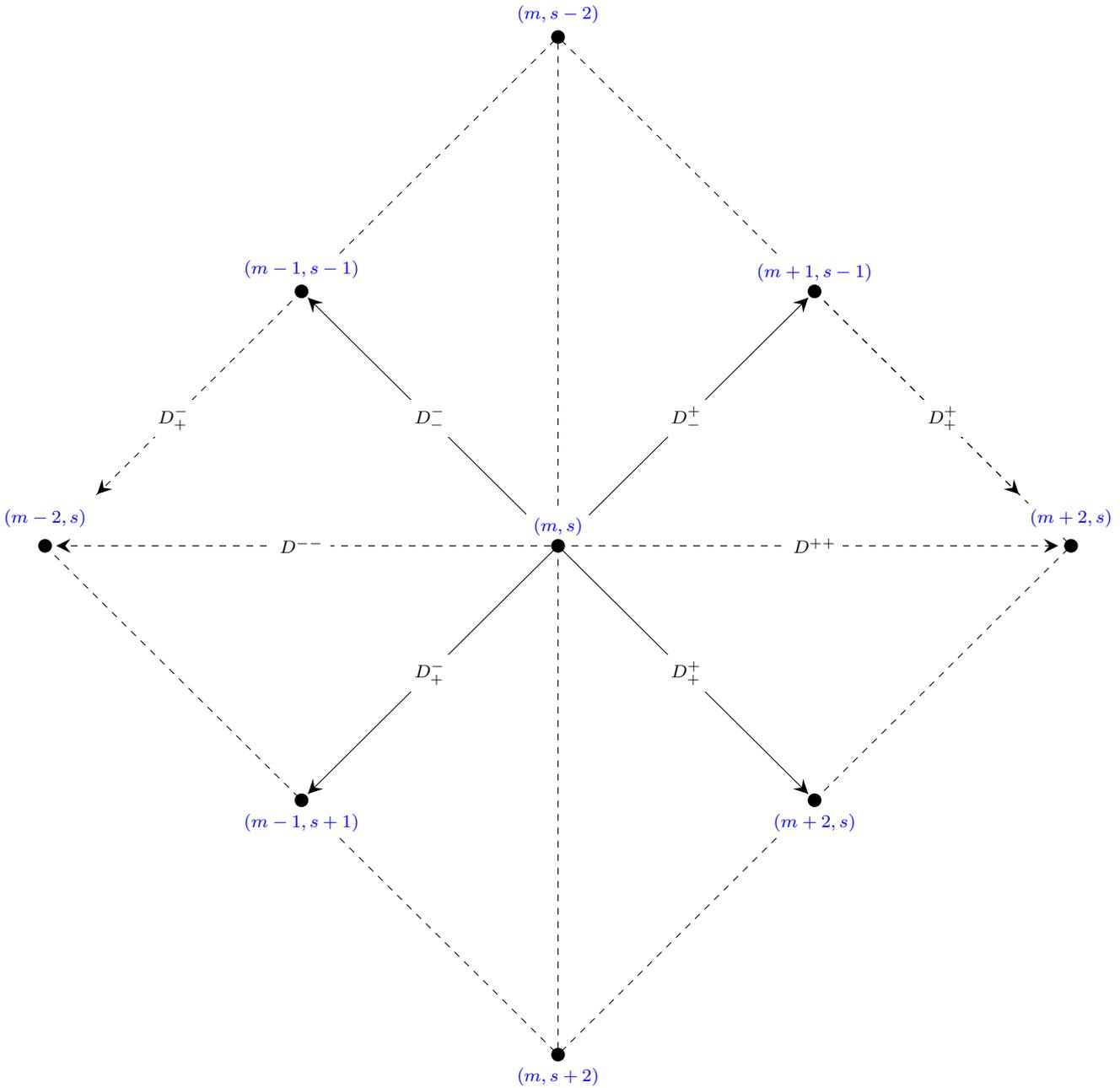

We define inverses with natural notation in the following way. Recalling that $D$ is the identity element we have
\begin{equation}
\label{invBaseOp}
D_-^-\circ D_+^+=D\quad\quad\rightarrow\quad\quad\,D_-^-\,\equiv\,\left(D_+^+\right)^{-1}\quad\quad\quad\textit{inverse},
\end{equation}
and similarely
\begin{equation}
D_-^+\circ D_+^-=D\quad\quad\rightarrow\quad\quad\,D_+^-\,\equiv\,\left(D_-^+\right)^{-1}\quad\quad\quad\textit{inverse}.
\end{equation}
This is a natural notation, as one thinks of the `$+$'s and `$-$'s as \textbf{canceling against each other} when they are \textit{on the same} script level\setcounter{footnote}{0}\footnote{so no subscript signs may cancel against superscript signs, or vice versa. Only superscript signs against superscript signs and subscript signs against subscript signs!}. Indeed the notation is right in suggesting that for instance 
\begin{equation}
D^{+++-+-\,-\,}=D^+\quad\quad\quad\textrm{and}\quad\quad\quad D^{-+-+-}_{+-+}=D^{-}_+.
\end{equation}
Also this structure is further explored in Appendix \ref{app:groups}. By application to $\gamma$, we find explicit expressions for the inverses as follows. By \eqref{BaseOp} and \eqref{invBaseOp} we find
\begin{equation}
\gamma=D\gamma=D_-^-\,D_+^+\gamma=D_-^-\,C_+^+\partial_c\gamma.
\end{equation}
Endoving the inverse $C_+^+$ with the same natural notation, $C_-^-\,\equiv\,(C_+^+)^{-1}$ (and similarly for $C_-^+$) we hence find the corresponding operator identities.
\begin{eqnarray}
\textit{Inverse basis-operators }\phantom{000000}
\begin{cases}
D_-^-=C_-^-\Box^{-1}\partial_c^*,\\
D_+^-=C_+^-\Box^{-1}\partial_c.
\end{cases}
\label{InvBaseOp}
\end{eqnarray}
Here we have invoked \eqref{Box}. The index form of $C_-^-$ and $C_+^-$ must be handled with care, however. The key lies in the fact that the inverse is taken at a different $(m,s)$-position, as a consequence of the fact that the coefficients are always evaluated at the end-position of each step. This is easy to see when we write it out, as follows.
\begin{eqnarray}
&(C_-^-)^m_s=\left((C_+^+)^{-1}\right))_s^m,\\
&(C_+^-)^m_s=\left((C_-^+)^{-1}\right))_s^m.
\end{eqnarray}
Comparing with \eqref{AandB}, and recognizing that the inverse in the last step is just the functional inverse, we necessarily find
\begin{equation}
(C_-^-)_s^m=2^{-\delta_{0s}}\frac{m+3+s}{(m+2)\chi}\quad\quad\quad\textrm{and}\quad\quad\quad (C_+^-)_s^m=2^{\delta_{0(s-1)}}\frac{m+3-s}{(m+2)\chi}.\label{AandB2}
\end{equation}
The inverses of the recursion relations \eqref{gammaRel} are now given as
\begin{eqnarray}
\label{InvgammaRel}
\textbf{Inverse rec. rel.s}\phantom{00000}
\begin{cases}
\gamma^{m-1}_{s-1}=D^-_-\gamma^m_s=(C_-^-)^{m-1}_{s-1}\partial_c\gamma^m_s\\
\gamma^{m-1}_{s+1}=D^+_-\gamma^m_s=(C_+^-)^{m-1}_{s+1}\partial_c^*\gamma^m_s.
\end{cases}
\end{eqnarray}
 Figure \ref{Fig1} illustrates the action of the basis-operators and their inverses.
\section{\quad Connection to existing literature}
\label{sec:Comp}
It is instructive at this point to compare with previous work on gravitational lensing. Kaiser $\&$ Squires described  a method for inverting the measured shear caused by the lens to obtain the surface mass density of a lens \cite{kaiser93}. To do so they started with the lensing equation. Their results are very well described by Schneider, who extended their work in two papers \cite{schneider95, schneider96}. Schneider finds a way to use only local data to reconstruct the lens-mass from the shear-field, and also points out the need for a generalized 'Kaiser-Squires' inversion procedrue, to account for stronger lensing effects. The standard procedure when starting from the lensing equation is to define the lensing potential 
\begin{equation}
\tilde{\psi}(\theta)=\frac{1}{\pi}\int_{\rm \mathbb{R}^2}{\rm d}^2\theta'\tilde{\kappa}(\bm{\theta'})\log\abs{\bm{\theta'-\theta}}.
\end{equation}
$\bm{\theta}$ is a vector of the two angles parametrizing the lensing 2-sphere, and $\tilde{\kappa}$ is the dimensionless surface mass density. We use a tilde ($\tilde{\,}$) to distinguish the notation from our own, since we have taken a different approach by starting from the GDE. Hence our lensing potential is defined differently \eqref{lensPot}. From the linearized lens mapping $\bm{\beta}=\bm{\theta}-\nabla\tilde{\psi}$ one defines now $\rm\textbf{d}\bm{\beta}=\bm{A}(\bm{\theta})\,\textbf{d}\bm{\theta}$, where $\bm{\beta}$ is the angular position on the source 2-sphere. Then
\begin{equation}
\label{A}
\bm{A}(\bm{\theta})=\left( {\begin{array}{cc}
1-\tilde{\kappa}+\tilde{\gamma}_1 & \tilde{\gamma}_2\\
\tilde{\gamma}_2 &1-\tilde{\kappa}-\tilde{\gamma}_1\\
\end{array} } \right),
\end{equation}
with
\begin{align}
&\label{1}\tilde{\kappa}(\theta)=\frac{1}{2}\nabla^2\tilde{\psi}(\theta),\\
&\label{2}\tilde{\gamma}_1=\frac{1}{2}(\tilde{\psi}_{,22}-\tilde{\psi}_{,11}),\\
&\label{3}\tilde{\gamma}_2=-\tilde{\psi}_{,12}.
\end{align}
Here $_{,i}$ refers to derivative w.r.t. $\theta_i$. From this, Kaiser $\&$ Squires obtained an inversion formula, yielding an expression for the surface mass density. Defining $\bm{\tilde{\gamma}}=\tilde{\gamma}_1+i\tilde{\gamma}_2$ one finds 
\begin{equation}
\tilde{\kappa}(\theta)=\frac{1}{\pi}\int_{\rm \mathbb{R}^2}{\rm d}^2\theta'\Re\left[\mathit{D}(\bm{\theta}-\bm{\theta}')\tilde{\gamma}(\bm{\theta'})\right],\quad\quad\textrm{(Kaiser-Squires inversion formula)}
\end{equation}
where $\mathit{D}=(\theta_1^2-\theta_2^2+2i\theta_1\theta_2)/\abs{\bm{\theta}}^4$. As mentioned, this formula gives the convergence (surface mass density) in terms of the shear-field generated by the lens.

\subsection{\quad  Connection to roulette amplitudes}
This section is just to show that we obtain similar expressions. Indeed, calculating the first few modes from \eqref{Gamma2} we find
\begin{align}
&\label{4}\alpha^1_0=-\chi^2\nabla^2\psi,\\
&\label{5}\alpha^1_2=-\frac{\chi^2}{2}(\psi_{\rm xx}-\psi_{\rm yy}),\\
&\label{6}\beta^1_2=-\chi^2\psi_{\rm xy}.
\end{align}
These equations have precisely the same structure as \eqref{1}-\eqref{3}\setcounter{footnote}{0}\footnote{Note that the lensing potential $\psi$ is in our case defined in Eq.~\eqref{lensPot}}. This means that the convergence-to-shear map that Kaiser $\&$ Squires have found, corresponds to the first few terms in the series expansion of roulette amplitudes. The higher order terms may in the same way be obtained from \eqref{Gamma2}. For an explicit example where the mass (convergence) and the derivatives of the mass distribution is calculated, please refer to \cite[Sec. IV]{clarkson16b}, where a circularly symmetric lens in the weak-field and thin-lens approximations is considered. In that work, the expressions \eqref{4}-\eqref{6} were inverted in a Kaiser-Squires like fashion to obtain the mass of the lens (Eqs. 149-154 therein) and its derivatives in terms of the roulette modes. In this work, we do the same (but more generally), except we restrict attention to expressing the derivatives of the potential (and not the mass and its derivatives) in terms of the roulette modes. This is done in the next section.

Before that however, a comment as to the relation between the variables of Eqs. \eqref{4}-\eqref{6} and Eqs. \eqref{1}-\eqref{3} is in place. Following \cite[Eq. 104]{clarkson16b}, one defines in the weak-field approximation with the Roulette expansion the amplification matrix $\mathcal{\bm{A}}$
\begin{equation}
\mathcal{\bm{A}}=\delta_{AB}-\chi^2\nabla_A\nabla_B\psi,
\end{equation}
where, recall, the potential is given by Eq. \eqref{lensPot}). From the above expression we find that $\mathcal{\bm{A}}$ is the same\footnote{on the level where one views $\mathcal{\bm{A}}$ and $\bm{A}$ as operators acting on defined potentials $\psi$ and $\tilde{\psi}$, respectively.} as $\bm{A}$ in Eq. \eqref{A}, and hence $\alpha^1_0=-\chi^2\tilde{\kappa}$, $\alpha^1_2=\chi^2\tilde{\gamma_1}$ and $\beta^1_2=\chi^2\tilde{\gamma_2}$.

\section{\quad  Derivatives of the potential in terms of roulette modes}
\label{sec:derivatives}
By equation \eqref{Alpha} and \eqref{Beta} we can, as mentioned, calculate all the higher order modes. 
To illustrate, we give in the following all the derivatives of the potential up to 4th order in terms of the roulette modes.

\begin{equation}
\begin{split}
&\psi_{\textrm{x}}=-\frac{1}{\chi}\alpha^0_1\quad,\quad\psi_{\textrm{y}}=-\frac{1}{\chi}\beta^0_1,\\
&\psi_{\textrm{xx}}\,=\, -\frac{1}{\chi ^2}\left(\alpha^1_0+ \alpha^1_2\right)\quad,\quad\psi_{\textrm{yy}}\,=\, -\frac{1}{ \chi^2}\left(\alpha^1_0-\alpha^1_2\right)\quad,\quad\psi_\textrm{xy}=-\frac{1}{\chi^2}\beta^1_2,\\
&\psi_{\textrm{3x}}\,=\, -\frac{1}{\chi ^3}\left(\alpha ^2_{1}+\alpha ^2_{3}\right)\quad,\quad\psi_{\textrm{3y}}\,=\,-\frac{1}{\chi ^3}\left(\beta^2_{1}-\beta ^2_{3}\right),\\
&\psi_{\textrm{xyy}}\,=\, -\frac{1}{3 \chi ^3}\left(\alpha ^2_{1}-3 \alpha ^2_{3}\right)\quad,\quad\psi _{\textrm{yxx}}\,=\, -\frac{1}{3 \chi ^3}\left(\beta^2 _{1}+3 \beta ^2_{3}\right),\\
&\psi _{\textrm{4x}}\,=\,-\frac{1}{\chi ^4}\left(\alpha ^3_{0}+ \alpha ^3_{2}+\alpha ^3_{4}\right)\quad,\quad\psi _{\textrm{4y}}\,=\,-\frac{1}{\chi ^4}\left(\alpha ^3_{0}- \alpha ^3_{2}+\alpha ^3_{4}\right),\\
&\psi _{\textrm{2x2y}}\,=\,-\frac{1}{3 \chi ^4}\left(\alpha ^3_{0}-3 \alpha^3 _{4}\right)\,,\,\psi _{\textrm{y3x}}\,=\,-\frac{1}{2 \chi ^4}\left(\beta ^3_{2}+2 \beta^3_{4}\right),
\psi _{\textrm{x3y}}\,=\,-\frac{1}{2 \chi ^4}\left(\beta^3_{2}-2 \beta^3_{4}\right),\\
&\psi _{\textrm{5x}} \,=\,-\frac{1}{\chi ^5}\left(\alpha ^4_{1}+\alpha ^4_{3}+\alpha ^4_{5}\right)\quad,\quad\psi _{\textrm{5y}}\,=\, -\frac{1}{\chi ^5}\left(\beta ^4_{1}-\beta^4_{3}+\beta^4 _{5}\right),\\
&\psi _{\textrm{4xy}}\,=\, -\frac{1}{5 \chi ^5}\left(\beta ^4_{1}+3 \beta ^4_{3}+5 \beta^4 _{5}\right)\quad,\quad\psi _{\textrm{x4y}}\,=\,-\frac{1}{5 \chi ^5}\left(\alpha^4_{1}-3 \alpha^4_{3}+5 \alpha^4_{5}\right),\\
&\psi_{\textrm{3xyy}}\,=\,\frac{1}{5 \chi ^5}\left(-\alpha^4_{1}+\alpha^4_{3}+5 \alpha^4_{5}\right)\quad,\quad\psi_{\textrm{xx3y}}\,=\,-\frac{1}{5 \chi^5}\left(\beta^4_{1}+\beta^4_{3}-5\beta^4_{5}\right)\quad,\\
&\textrm{etc. }.
\end{split}
\end{equation}
For brevity we used, in the above, notation such that $\partial_\textrm{x}^n\partial_\textrm{y}^m\psi\,\equiv\,\psi_{n\textrm{x}m\textrm{y}}$, where $n,m$ are positive integers.

\section{\quad Conclusion}
\label{sec:concl}
In this work we have generalized the Kaiser-Squires relations to strong lenses. This was done deriving a set of recursion relations for gravitational lensing (in the flat-sky, weak-field and thin-lens approximations) using the Roulette formalism. We have presented a simplified version of this formalism by formulating the theory in complex variables, and by introducing $D$-operators. We have found a very simple expression for the roulette amplitudes, invoking no sums or integrals \eqref{GamNice1}. This is beneficial for computational purposes. 

We expect that recursion relations similar to those found in this work, will also hold true in more general implementations of the Roulette expansions. One should therefore investigate if the same structures are seen in the absence of the flat-sky, weak-field and thin lens approximations used here. This, however, is left for future work.

\newpage
\begin{appendix}
\section{\quad Proof of recursion relations}
\label{App:A}
In this appendix we derive the recursion relations \eqref{gammaRel},
\begin{equation}
\gamma_{s+1}^{m+1}=(C_+^+)^{m+1}_{s+1}\partial_{\rm c}\gamma_s^m\quad\quad\textrm{and}\quad\quad\gamma_{s-1}^{m+1}=(C_-^+)^{m+1}_{s-1}\partial_{\rm c}^*\gamma_s^m.
\end{equation}
Take first the definition of $\gamma_{s+1}^{m+1}$, equation \eqref{Gamma2}. According to it one finds that the term with $\partial_{\rm x}^{m-k+1}\partial_{\rm y}^k\psi$ is given by 
\begin{equation}
\label{Aeq1}
-2^{-\delta_{0s}}\chi^{m+2}{m+2\choose k+1}\mathcal{D}_{s+1}^{m+1(k+1)}\partial_{\rm x}^{m-k+1}\partial_{\rm y}^k\psi.
\end{equation}
Similarly, starting from the recursion relation $\gamma_{s+1}^{m+1}=(C_+^+)^{m+1}_{s+1}\partial_{\rm c}$ instead, one finds the corresponding terms
\begin{equation}
\label{Aeq2}
-\chi^{m+1}(C_+^+)_{s+1}^{m+1}\left({m+1\choose k+1}\mathcal{D}_s^{m(k+1)}+i{m+1\choose k}\mathcal{D}_s^{m(k)}\right).
\end{equation}
Equating equation \eqref{Aeq1} with \eqref{Aeq2} and inserting for $(C_+^+)^{m+1}_{s+1}$ gives, after some straight forward algebraic manipulation, the equation 
\begin{equation}
\label{Rel1}
(m+3+s)\mathcal{D}_{s+1}^{m+1(k+1)} =(m+1-k)(\mathcal{D}_{s}^{m(k+1)})+i(k+1)\mathcal{D}_{s}^{m(k)}.\\
\end{equation}
Following the same procedure for $\gamma_{s-1}^{m+1}$ similarely gives the equation
\begin{equation}
\label{Rel2}
(m+3-s)\mathcal{D}_{s-1}^{m+1(k+1)} =(m+1-k)(\mathcal{D}_{s}^{m(k+1)})+i(k+1)\mathcal{D}_{s}^{m(k)}.\\
\end{equation}
The two relations \eqref{Rel1} and \eqref{Rel2} must hold for the recursion relations to be correct. In Appendix \ref{App:Trig} they are shown to hold.

\section{\quad Trigonometric relations}
\label{App:Trig}
Invoking the manipulation ${\rm e}^{i\cdot(s+1)\theta}={\rm e}^{i\cdot s\theta}\cdot(\cos\theta+i\sin\theta)$ in equation \eqref{D} immediately gives
\begin{eqnarray}
\label{manip1}
\mathcal{D}_{s+1}^{m(k)}=\mathcal{D}_s^{m+1(k)}+i\mathcal{D}_s^{m+1(k+1)}.
\end{eqnarray}
Invoking instead the manipulation $\cos\theta^2+\sin\theta^2=1$ in equation \eqref{D} gives
\begin{eqnarray}
\label{manip2}
\mathcal{D}_{s}^{m(k)}=\mathcal{D}_s^{m+2(k)}+\mathcal{D}_s^{m+2(k+2)}.
\end{eqnarray}
From differentiation w.r.t. $\theta$ (or similarly; from integration by parts), one finds 
\begin{eqnarray}
\label{DerRel}
\frac{{\rm d}}{{\rm d}\theta}\mathcal{D}_s^{m(k)}=0\quad\quad\rightarrow\quad\quad s\mathcal{D}_s^{m(k)}=-(m+1-k)\mathcal{D}_s^{m(k+1)}+ik\mathcal{D}_s^{m(k-1)}.
\end{eqnarray}
\subsection{\quad  Proof of relations \eqref{Rel1} and \eqref{Rel2}}
Add (i) first $(m+1)\mathcal{D}_s^{m(k)}$ to both sides of the above equation, (ii)  let thereafter $\{m,k,s\}\rightarrow\{m+1,k+1,s+1\}$ and (iii) use the manipulations \eqref{manip1} and \eqref{manip2} to obtain
\begin{eqnarray}
\label{ImpRel1}
& (m+3+s)\mathcal{D}_{s+1}^{m+1(k+1)}=(m+1-k)(\mathcal{D}_{s}^{m(k+1)})+i(k+1)\mathcal{D}_{s}^{m(k)}.
\end{eqnarray}
Following the same procedure as above, but now by (i)  \textit{subtracting} first $(m+1)\mathcal{D}_s^{m(k)}$ on both sides of equation \eqref{DerRel}, (ii) thereafter letting $\{m,k,s\}\rightarrow\{m+1,k+1,s-1\}$ and finally (iii) using the manipulations \eqref{manip1} and \eqref{manip2}, one ends up with
\begin{equation}
\label{ImpRel2}
(m+3-s)\mathcal{D}_{s-1}^{m+1(k+1)} =(m+1-k)(\mathcal{D}_{s}^{m(k+1)})+i(k+1)\mathcal{D}_{s}^{m(k)}.\\
\end{equation}
Thus the relations \eqref{Rel1} and \eqref{Rel2} hold.
\subsection{\quad  Additional relations}
One may also note relations like
\begin{align}
&\mathcal{D}_s^{m(k)}=\mathcal{D}_s^{m-2\,(k)}-\mathcal{D}_s^{m(k+2)},\\
&\mathcal{D}_s^{m+1\,(k)}=\frac{1}{2}\left(\mathcal{D}_{s-1}^{m(k)}+\mathcal{D}_{s+1}^{m(k)}\right),
\end{align}
which follow from manipulation with trigonometric identities. In this paper we omit the (straight forward) proofs, as we make no use of these relations. Take now the definition
\begin{eqnarray}
&\mathcal{D}_s^{m(k)}=\mathcal{C}_s^{m(k)}+i\mathcal{S}_s^{m(k)}=\frac{1}{\pi}\int_{-\pi}^{\pi}{\rm d}\theta\sin^k\theta\cos^{m-k+1}\theta\,{\rm e}^{i\cdot s\theta}\label{D},
\end{eqnarray}
and recall that
\begin{equation}
\label{binomial}
\left(a+b\right)^n=\sum_{k=0}^{n}{n\choose k}a^{n-k}\left(ib\right)^{k}.
\end{equation}
Then
\begin{align*}
\begin{split}
&\mathcal{D}_{s+n}^{m(k)}=\mathcal{C}_{s+n}^{m(k)}+i\mathcal{S}_{s+n}^{m(k)}=\frac{1}{\pi}\int_{-\pi}^{\pi}{\rm d}\theta\sin^k\theta\cos^{m-k+1}\theta\,{\rm e}^{i\cdot s\theta}(\cos\theta+i\sin\theta)^n\\
&=\frac{1}{\pi}\int_{-\pi}^{\pi}{\rm d}\theta\sin^k\theta\cos^{m-k+1}\theta\,{\rm e}^{i\cdot s\theta}\sum_{k=0}^{n}{n\choose k}\cos^{n-k}\left(i\sin\right)^{k}\\
&=\sum_{j=0}^{n}(i)^{j}{n\choose j}\frac{1}{\pi}\int_{-\pi}^{\pi}{\rm d}\theta\sin^k\theta\cos^{m-k+1}\theta\,{\rm e}^{i\cdot s\theta}\cos^{n-j}\left(\sin\right)^{j}\\
&=\sum_{j=0}^{n}(i)^{j}{n\choose j}\frac{1}{\pi}\int_{-\pi}^{\pi}{\rm d}\theta\sin^{k+j}\theta\cos^{m+n-(k+j)+1}\theta\,{\rm e}^{i\cdot s\theta}\\
&=\sum_{j=0}^{n}(i)^{j}{n\choose j}\mathcal{D}_{s}^{m+n(k+j)}
\label{D}.
\end{split}
\end{align*}
We therefore have the generic relation
\begin{equation}
\mathcal{D}_{s+n}^{m(k)}=\sum_{j=0}^{n}(i)^{j}{n\choose j}\mathcal{D}_{s}^{m+n(k+j)}.
\end{equation}
\newpage

\section{\quad $D$-operators as a group}
\label{app:groups}
The $D$-operators have a very intuitive notation, and hence they are very easy to use. Now that we have the basis-operators $D_-^+$ and $D_+^+$ in store, we can construct any other $D$-operator by extending our formalism in the most natural of manners. 

\begin{dfn}[Cancellation of signs]
Let $b,c,d,e\,\in\,\mathbb{Z}$, and denote with $b+$ the sequence of $b$ positive signs ($+$) if $b$ is positive and the sequence of $b$ negative ($-$) signs if $b$ is negative. Define now the operation $\circ$ such that
$$D_{c+}^{b+}\circ D_{e+}^{d+}=D_{(c+e)+}^{(b+d)+}.$$
\end{dfn}
With this definition the notation is endowed with an intuitive `cancellation of signs' property. For instance $D_{+++-+}^{+++-++}=D_{+++}^{++++}=D_{3+}^{4+}$. It is also straigth forward to show that the $D$-operators commute as operators applied to $\gamma$. For the basis-operators we find
\begin{equation}
\left[D_-^+,D_+^+\right]\gamma=(D_-^+\circ\,D_+^+-D_+^+\circ\,D_-^+)\gamma=(D^{++}-D^{++})\gamma=\gamma^{++}-\gamma^{++}=0 
\end{equation}
as expected. Note now some further properties. First of all, since $D^{\rm *}\gamma=D_{\rm *}\gamma=0$ (where ${\rm *}\,\in\,\{+,-\}$), and since \textit{any} operator $D_{b+}^{a+}$ can be decomposed into the two basis-operators $D_-^+$ and $D_+^+$ (and their inverses), alongside $D_{\rm *}$, we can conclude that \textit{any} operator with an odd total number of signs will vanish. 

\begin{dfn}[Set of proper D operators]
Consider the set $\tilde{D}$ of all operators with an even total number of signs, as defined over allowed $(m,s)$-values. From the definition of $C_+^+$ and $C_-^+$ (with corresponding inverses) it is clear that all of these are non-zero. One must have that
\begin{equation}
D_{b+}^{a+},\in\,\tilde{D}\quad\quad\iff \quad\quad D_{b+}^{a+}=(D_+^+)^n\circ (D_-^+)^m
\end{equation}
for $a,b,n,m\,\in\,\mathbb{Z}$.
\end{dfn}
Straight forward algebraic manipulation now shows that $n=(a+b)/2$ and $m=(a-b)/2$.

\paragraph{Group structure:} Finally; note that $(\tilde{D},\circ)$ is a group.
\begin{proof}
Take $a,b,c,d,e,f\,\in\,\mathbb{Z}$ and let $D^{a+}_{b+}$, $D^{c+}_{d+}$ and $D^{c+}_{d+}$ be arbitrary members of the group. Then (i) there is an identity element $D$, since $D\circ D^{a+}_{b+}= D^{a+}_{b+}\circ D=D^{a+}_{b+}$,
\linebreak (ii) closure is fulfilled, since $D^{a+}_{b+}\circ D^{c+}_{d+}=D^{(a+c)+}_{(b+d)+}$ is also a member of the group, \linebreak (iii) associativity is fulfilled, since $D^{a+}_{b+}\circ (D^{c+}_{d+}\circ D^{e+}_{f+})=D^{(a+c+e)+}_{(b+d+f)+}=(D^{a+}_{b+}\circ D^{c+}_{d+})\circ D^{e+}_{f+}$ and (iv) there is an inverse $D^{a-}_{b-}$ for every member $D^{a+}_{b+}$.
\end{proof}

\newpage
\section{\quad Relating $\gamma_{s'}^{m'}$ to $\gamma_s^m$}
\label{App:RelatingGammas}
This appendix builds on the forgoing one. Here we give a receipt for relating any two $\gamma$s to each other.
\begin{enumerate}
\item Ensure that the $(m,s)$-position has a non-zero $\gamma$ and denote it as $\gamma^m_s$. 
\item Find the differences $M=m'-m$ and $S=s'-s$. The $D$-operator equation (the recursion relation from $\gamma_{s}^{m}$ to $\gamma_{s'}^{m'}$) is then
\begin{equation}
\gamma^{M\,+}_{S\,+}=D^{M\,+}_{S\,+}\gamma,
\end{equation}
with notation as defined in the previous appendix. Hence $M,S$ can also be negative. For instance, if the difference $S$ is negative, then $S+\,=\,(-\abs{S})+\,=\,\abs{S}-$.
\item If $M+S$ is odd, then $D$ is according to the previous appendix not proper. Since we start with non-zero $\gamma$ we must then find $\gamma^{M\,+}_{S\,+}=0.$ Otherwise (refer to the forgoing appendix) we must have
\begin{eqnarray}
D_{S+}^{M+}=\left(D_+^+\right)^{\frac{M+S}{2}}\left(D_-^+\right)^{\frac{M-S}{2}}.
\end{eqnarray}
\item By \eqref{BaseOp} one now finds for $\gamma_{s'}^{m'}$ the following.

\begin{eqnarray}
\gamma_{s'}^{m'}&=\gamma_{s+S}^{m+M}=\left(D_+^+\right)^{\frac{M+S}{2}}\left(D_-^+\right)^{\frac{M-S}{2}}\gamma_s^m.\\
\end{eqnarray}
Note that $(M-S)/2$ is always a natural number. 
\item From here one may write down a coordinate expression. There will, however, be many paths connecting two $\gamma$s, and the expression will be path dependent. One must make sure to choose a path that contains only $(m,s)$-positions where $\gamma$ is defined. Also remember to use the inverses as basis-operators if the exponents are negative\setcounter{footnote}{0}\footnote{If for instance $(M+S)/2$ is negative, then use rather $C_-^-$ as a base, so that the exponent becomes positive (in this case $\abs{(M+S)/2}$).}. With these precautions, the rest is straight forward. Here we suffice it to give an expression for the case where both exponents are non-negative. In order to ensure that we only go through valid $(m,s)$- positions we move along $D_+^+$ first.
\begin{equation}
\label{long}
\gamma^{m+M}_{s+S}=\left(\prod_{i=1}^{A^-}\left(C_-^+\right)_{s+A^+-i}^{m+A^++i}\prod_{j=1}^{A^+}\left(C_+^+\right)_{s+j}^{m+j}\right)\Box^{A^-}\partial_c^S\gamma_s^m,
\end{equation}
where $A^\pm\,=\,(M\pm S)/2$ for brevity. Note the product sums, which follow as a consequence of iterating over successive $(m,s)$ values as we move along. Very similar expressions will be found also for the cases where the exponents are negative. These expressions relate the $\gamma$s at any two valid positions in the $(m,s)$-configuration-space to each other. 

\end{enumerate}
\newpage
\section{\quad Derivation of the simpler expression for $\gamma_s^m$}
\label{App:DerOfPot}
From the last appendix, and recalling equation \eqref{lowestmode}, it is evident that in order to express the modes in terms of the potential, all we need to do is to  insert $m=0$ and $s=1$ in the last equation. Starting at the lowest rung, $\gamma_0^1$, we will not need the inverses. It is  cleanest now to use only the tuple $(m,s)$ as variables. Letting therefore $M\,\rightarrow\,m$ and $S\,\rightarrow\,s-1$, equation \eqref{long} becomes
\begin{equation}
\label{e}
\gamma^{m}_{s}=-\chi\left(\prod_{i=1}^{a^-}(C^+_-)_{1+a^+-i}^{a^++i}\prod_{j=1}^{a^+-1}(C^+_+)_{j+1}^{j}\right)\Box^{a^-}\partial_c^s\psi,
\end{equation}
where we have also inserted $\gamma^0_1=-\chi\partial_{\rm c}\psi$, and defined
\begin{equation}
a^{\pm}=\frac{m+1\pm s}{2},\quad\textrm{which implies}\quad m+1=a^++a^- \quad \textrm{and}\quad s=a^+-a^-.
\end{equation}
One may show that
\begin{equation}
\prod_{j=1}^{a^+-1}(C^+_+)_{j+1}^{j}=\left(\frac{\chi}{2}\right)^{a^+-1}\quad\quad\quad\quad\textrm{and}\quad\quad\quad\quad \prod_{i=1}^{a^-}(C^+_-)_{1+a^+-i}^{a^++i}=\left(\frac{\chi}{2}\right)^{a^-}\prod_{i=1}^{a^-}\frac{a^++i}{i}.
\end{equation}
Using now
\begin{equation}
{ n \choose k }=\prod_{i=1}^k\frac{n+1-i}{i}\,=\,\prod_{i=1}^k\frac{n-k+i}{i}
\end{equation}
one finds\setcounter{footnote}{0}\footnote{Note that $\partial_c^{a^+}\partial_c^{*a^-}=\Box^{a^{-}}\partial_c^s$. Hence we may also write $\gamma_s^m=\Gamma_s^m\partial_c^{a^+}\partial_c^{*a^-}\psi.$ }
\begin{equation}
\label{AppGNice}
\gamma_s^m=\Gamma_s^m\Box^{a^{-}}\partial_c^s\psi,
\end{equation}
where
\begin{eqnarray}
\Gamma_s^m=
\begin{cases}
-(2^{-\delta_{0s}})\frac{\chi^{m+1}}{2^m}{m+1\choose (m+1-s)/2}\quad \quad\quad m+s\quad\textrm{odd,}\\
\quad\quad 0\quad\quad\quad\quad\quad\quad\quad\quad\textrm{else}.
\end{cases}
\end{eqnarray}
This expression is particularely useful for  numerical purposes. In contrary to Equation \eqref{e}, this expression also ensures that $s\,\leq\,m+1$ (due to the binomial), which is as it should. Note that we also require $s\,\geq\,0$.

By use of the identities in  \ref{App:NoteOnDerivatives}, we now find 
\begin{align}
&\alpha_s^m=\Gamma_s^m\Box^{a^-}\sum_{k=0}^{2k\,\leq\,s}(-1)^k{s \choose 2k}\partial_{\rm X}^{s-2k}\partial_{\rm Y}^{2k},\\
&\beta_s^m=\Gamma_s^m\Box^{a^-}\sum_{k=0}^{2k+1\,\leq\,s}(-1)^k{s \choose 2k+1}\partial_{\rm X}^{s-2k-1}\partial_{\rm Y}^{2k+1}.
\end{align} 

\section{\quad A note on $\partial_c^n$}
\label{App:NoteOnDerivatives}
Everything in the above expression is straight forward, except for the factor $\partial_c^S$. In splitting $\gamma$ into $\alpha$ and $\beta$ it is useful to invoke the binomial formula with complex numbers,
\begin{equation}
\label{binomial}
\left(\partial_X+i\partial_Y\right)^n=\sum_{k=0}^{n}{n\choose k}\partial_X^{n-k}\left(i\partial_Y\right)^{k}.
\end{equation}
The even and odd terms now give the real and imaginary parts, respectively. We have 
\begin{align}
&\Re\{\partial_c^n\}=\Re\{\left(\partial_X+i\partial_Y\right)^n\}=\sum_{k=0}^{2k\,\leq\,n}{n\choose 2k}\partial_X^{n-2k}(-1)^k\partial_Y^{2k},\\
&\Im\{\partial_c^n\}=\Im\{\left(\partial_X+i\partial_Y\right)^n\}=\sum_{k=0}^{2k+1\,\leq\,n}{n\choose 2k+1}\partial_X^{n-2k-1}(-1)^k\partial_Y^{2k+1}.
\end{align}
Hence we can always isolate the real and imaginary parts. Also note that 
\begin{equation}
\label{binomialBox}
\Box^n=\left(\partial^2_X+\partial^2_Y\right)^n=\sum_{k=0}^{n}{n\choose k}\partial_X^{2(n-k)}\partial_Y^{2k}.\\
\end{equation} 

\end{appendix}
\end{document}